\documentclass[sigconf]{acmart}
\usepackage[utf8]{inputenc}
\usepackage[T1]{fontenc}
\usepackage[english]{babel}
\usepackage{blindtext}

\renewcommand\footnotetextcopyrightpermission[1]{} 
\setcopyright{none}

\acmDOI{}

\acmISBN{}

\acmPrice{}

\begin{document}
\title{A more secure IPv6 neighborhood process}


\author{Camilla Jacome}
\affiliation{
    \institution{IFPB}
    \city{João Pessoa} 
    \state{Para\'iba}
}
\email{camilla.jacome@academico.ifpb.edu.br}

\author{Michael Monteiro}
\affiliation{
    \institution{IFPB}
    \city{João Pessoa} 
    \state{Para\'iba}
}
\email{michael.monteiro@academico.ifpb.edu.br}

\author{Miguel Cabral}
\affiliation{
    \institution{IFPB}
    \city{João Pessoa} 
    \state{Para\'iba}
}
\email{miguel.cabral@academico.ifpb.edu.br}

\author{Pedro Carvalho}
\affiliation{
    \institution{IFPB}
    \city{Patos} 
    \state{Para\'iba}
}
\email{pedro.carvalho@ifpb.edu.br}

\author{Leandro Almeida}
\affiliation{
    \institution{IFPB}
    \city{João Pessoa}  
    \state{Para\'iba}
}
\email{leandro.almeida@ifpb.edu.br}
\renewcommand{\shortauthors}{Jacome C.et al.}
\begin{abstract}
The process of neighborhood establishment in an IPv6 network is made out through the NDP (Neighbor Discovery Protocol). Using ICMPv6 messages (NS - Neighbor Solicitation and NA - Neighbor Advertisiment) which contains the IP address to be resolved, is exposed during the exchange of messages, making communication vulnerable to various types of attacks. This paper presents a proposal for a safer neighborhood establishment using the DH and HMAC algorithms. Experiments were performed in virtualized environments with the objective of analyzing the efficiency of the modification proposed in the NDP.
\end{abstract}
\maketitle
\section{Introduction}
The NDP assists IPv6 in several operations on a local network, such as: neighbors discovery, duplicate addresses detection, routers discovery, reachability maintenance and physical addresses resolution. Is assumed that all nodes in an IPv6 network are trusted, becoming a target for many attacks on computer networks \cite {rfc3756}, such as: Spoofing, Denial of Service, Replay attack, Redirection attack and rogue router attack \cite {6148204}. Over the years some solutions have been proposed, among the most promising ones, the SeND (Secure Neighbor Discovery). However, it demands high consumption of computational resources, thus occasionally limiting the use scope of SeND \cite {8022867}.

This work proposes an improvement to the IPv6 and NDP protocols, in order to allow a more secure adjacency establishment process using HMAC (Hash-Base Message Authentication Code) in conjunction with the DH (Diffie-Hellman) algorithm \cite {1976-diffie}.

\section{Proposal}
With the use of HMAC in conjunction with DH, the solution proposes a hash function to be applied in the ICMPv6’s header target field in an NS message. The Target field contains the IPv6 address to be resolved. Hiding the hashed IP, it is assumed that if the NS message is intercepted by an attacker who has not established a secure channel on the network with the host previously through DH, the attacker can not falsify the NA response for NS, as it will not know which IP address is contained in the Target field.

First, there is the key establishment process using DH so that all hosts on the network have a secret shared between them. When it is necessary for one host to create a connection with another to request the MAC address and establish neighborhood, the host sends an NS message in multicast-all-nodes in the network. HMAC is used as a hash function on the IP address contained in the Target field using the previously established key between the hosts.

Upon receiving the NS, the hosts will compare the pre-calculated hash of their IP address with the hash contained in the Target of the received packet. However, only the true host will authenticate, because it will be the only one that hash will be equal and the attacker would not know which IP address to forge because it would have to solve the hash contained in NS, which would give advantageous time to the true host of the NS message reply with the NA message containing it’s MAC address.

\subsection{Proposal weaknesses}
Despite increasing the security level by hiding the IP address during the neighborhood establishment process, the proposal has some weaknesses, among them:

\begin{itemize}
\item How the proposal focuses on the NS/NA process, does not protect against attacks after this message exchange, besides depending on the DH procedure do not suffer MitM.
\item For the reason of generating a pair of keys between all hosts on the network, in a large network where a large number of hosts can join at the same time, this generation of keys between them can generate an overhead.
\end{itemize}

\section{Results}
    
Three virtual machines have been created, all of which are Linux Debian version 9.6. Was used VMWare ESXi in a server with eight cores, 32 GB RAM and 1 TeraByte HD. The idea was that Alice wants to establish neighborhood with Bob and the Intruder is an attacker in the network who tries to pass himself as Bob and falsify the answer of Alice\textquotesingle s NS with a false NA.

Our proposal\textquotesingle s script needs the following dependencies: python3, python3-pip, scapy3, and pyDHE. The first part of the experiments was to know the difference in the use of computational resources between the normal NDP process and the NDP process of our proposal. It was used the Collectl tool in Alice\textquotesingle s machine to collect and store performance data from computers and network equipment. The script was run 30 times for both the normal NDP process and the NDP process of our proposal, each time CPU, memory and network data were collected. After obtaining the data, the average of the 30 executions was made. The mean of the results is shown in Figure 1.

\begin{figure}[!htb]
        \includegraphics[width=8cm]{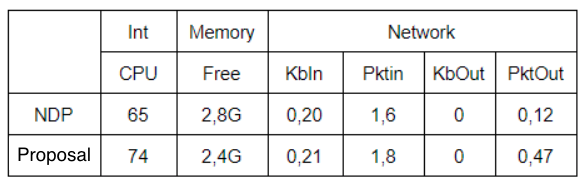}
        \caption{Use of Computational Resources}
        \label{Rotulo}
    \end{figure}

In figure 1 it can be seen that in the proposal CPU usage was only 9\% higher. In memory we can see that the proposal\textquotesingle s script consumes 400M more and about the network we can see that the biggest difference is in the incoming and outgoing packets, where the proposal consumes almost four times more of the outgoing packets, it is supposed that this difference is due to the DH key exchange at the beginning of the process. This key exchange can also be the explanation for the increased CPU and memory consumption.

The second part of the experiments was carried out to test the level of security of the proposal. In the script there is a function that was created to perform the Man-in-the-Middle attack. This function monitors network traffic and, when capturing an ICMPv6 message of type 135 (an NS), it responds with several NA messages passing through the target host. The normal NDP process was performed, where the Intruder pretends to be Bob and Alice accepts this as true. To view traffic, the Wireshark tool was used, which is a program that analyzes network traffic.

In contrast, when the same procedure was performed using the applied proposal, the Intruder did not succeed in pretending to be Bob, since he could not see the IP address of Bob that was contained in the NS Target field sent by Alice. You can see in Figure 2 in the traffic captured by Wireshark, the NS of the proposal sent by Alice in multicast and the NA of Bob sent in unicast to Alice. It is also noted that the Target field does not contain Bob's IPv6 address, but it\textquotesingle s hash.

 \begin{figure}[!htb]
        \includegraphics[width=9.5cm]{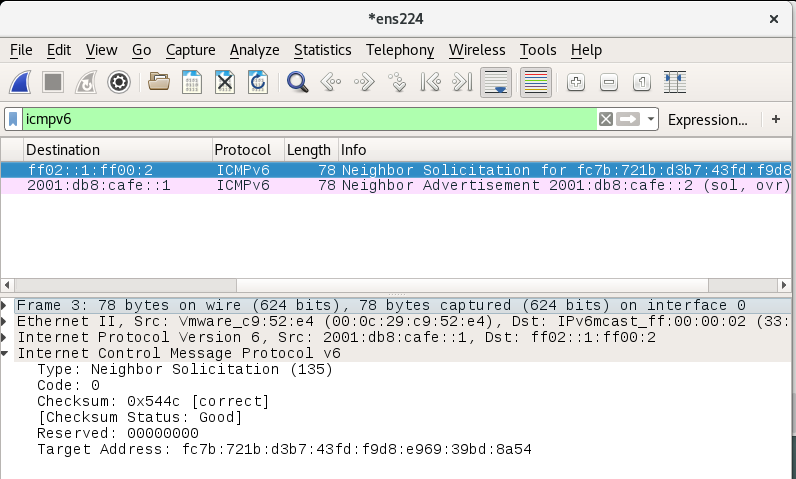}
        \caption{Print Screen NS/NA}
        \label{Rotulo}
    \end{figure}

Considering the experiments carried out, it is conjectured that the proposal obtains a higher level of security compared to the normal NDP process.

\section{Conclusion}
In the midst of the current situation in which we live in which technology is becoming more and more present in society and with the emergence of the Internet of Things, IPv6 has become a reality.

In the literature, many works point out problems related to the process of neighborhood discovery of IPv6 in conjunction with the NDP. In view of this problem, this work is a more secure example of neighborhood establishment in IPv6 networks. Hiding the IP during the Neighborhood Discovery, although it does not solve all the problems related to the NDP, increases the security level of the process and hinders some attacks, as demonstrated in the experiments, meeting the premise of the proposal.

In future works will be used Software Defined Networking (SDN) to solve the weaknesses presented. The idea is a virtual switch for key management, that is, all hosts will have a key with the switch before establishing the Diffie-Hellman key exchange, thus protecting against the Man-in-the-Middle attack during this process. To avoid network overhead, a virtual switch can allocate more resources when needed.
\bibliographystyle{ACM-Reference-Format}
\bibliography{paper}
\end{document}